\documentclass[aip,apl,amsmath,amssymb,preprint,reprint]{revtex4-1}
\usepackage[dvipsnames]{xcolor}
\usepackage{siunitx}
\usepackage{graphicx}
\usepackage{dcolumn}
\usepackage{bm}

\usepackage{ulem}

\usepackage[utf8]{inputenc}
\usepackage[T1]{fontenc}
\usepackage{mathptmx}
\usepackage{etoolbox}

\newcommand{\om}[1]{\omega_\textrm{#1}}
\newcommand{\Om}[1]{\Omega_\textrm{#1}}

\makeatletter
\def\@email#1#2{%
 \endgroup
 \patchcmd{\titleblock@produce}
  {\frontmatter@RRAPformat}
  {\frontmatter@RRAPformat{\produce@RRAP{*#1\href{mailto:#2}{#2}}}\frontmatter@RRAPformat}
  {}{}
}%
\makeatother
\begin{document}

\preprint{AIP/123-QED}

\title{Acoustic damping limits superfluid helium as a thermal bath for mechanical resonators}
\author{Y. Huang}\email{yunhu@ualberta.ca}
\author{M.\ Ebrahimi}
\author{A.\ Rashedi}
\author{J.\ P.\ Davis}%
\affiliation{Physics Department, University of Alberta.}

\date{\today}

\begin{abstract}
Thermalization is a persistent problem for cryogenic mechanical resonators. The weak supports used to obtain high mechanical quality factors are also poor thermal links, while the optical or microwave fields used for readout and control can heat the resonator above the cryostat temperature. Superfluid helium is therefore an appealing exchange medium: it has excellent low-temperature thermal transport and an inviscid superfluid component, suggesting cooling of mechanical devices without the damping associated with an ordinary fluid. Here we test this idea by immersing a \qty{250}{\micro\meter}-diameter yttrium iron garnet sphere in liquid \textsuperscript{4}He and reading out a \(\sim\qty{12.47}{\mega\hertz}\) mechanical mode using cavity magnomechanics. The mechanical linewidth increases from \qty{6.1}{\hertz} in vacuum to \qty{9.1}{\kilo\hertz} after filling the cavity with liquid helium in the normal state, and reaches \qty{17.5}{\kilo\hertz} deep in the superfluid at \qty{84}{\milli\kelvin}. Thus the mechanical quality factor is reduced by more than three orders of magnitude, and crossing into the superfluid does not recover low dissipation. At the same time, the magnon linewidth remains strongly dependent on microwave power at the lowest temperatures, showing that the sphere is not perfectly anchored to the cryostat despite the large mechanical damping. The result follows from acoustic radiation: the surface motion of the sphere efficiently launches gapless longitudinal sound into the liquid, and a radiation-impedance estimate gives a linewidth of order \qty{25}{\kilo\hertz}. Superfluid helium is therefore not a low-loss thermal bath for mechanical modes that efficiently radiate compression waves. Shear motion, on the other hand, may provide a route around this limitation.
\end{abstract}

\maketitle


Thermalizing a mechanical resonator to the cryostat temperature is often difficult. The weak clamping used to obtain high mechanical quality factors also provides a poor thermal link, while optical or microwave dissipation can heat a device above the surrounding bath\cite{Aspelmeyer2014,OConnell2010,Teufel2011,Reinhardt2016,Norte2016,Tsaturyan2017,Fink2016,Ramp2019,MacCabe2020,Meenehan2014,Meenehan2015,MacDonald2016,Kim2016,Hauer2018}. Exchange gas helps near \qty{4}{\kelvin}, but this route is lost at millikelvin temperatures\cite{MacDonald2016,Kim2016}. Superfluid helium therefore appears attractive: it can surround a device without rigid mechanical contact, has excellent thermal transport, and its superfluid component is inviscid. Bradley \textit{et al.}~described a superfluid as an effective ``mechanical vacuum'' and proposed superfluid \textsuperscript{3}He as a thermal contact for MHz nanomechanical resonators\cite{Bradley2017}. Superfluid \textsuperscript{4}He has likewise improved thermalization in superconducting-circuit experiments\cite{Lane2020,Li2025}. This raises a simple question: can a superfluid act as an exchange medium for mechanics, providing strong thermal contact without adding mechanical loss?

There is already a large body of work placing mechanical resonators in both superfluid \textsuperscript{4}He and \textsuperscript{3}He. Historically, vibrating wires and quartz tuning forks have been used as thermometers and probes of quantum fluids, while more recent MEMS and NEMS devices have pushed these measurements to higher frequencies and smaller length scales\cite{Bradley2017,Guenault2019He4,Barquist2020,Guenault2019He3,Zheng2016,Zheng2017,Jiang2023}. In most of these experiments the resonator was used as a probe of the fluid rather than as a device whose thermalization was itself the question. A parallel body of work uses the helium itself as the mechanical degree of freedom, including nanofluidic Helmholtz modes and cavity optomechanics with first, third, fourth, and Brillouin sound modes\cite{Rojas2015,Souris2017,Baker2016,Kashkanova2017,Shkarin2019}. In those experiments the helium is the resonator, so they do not directly address the question here: whether helium can thermalize a separate high-Q solid resonator without strongly damping it.

However, low viscosity is not the same as weak mechanical coupling. A liquid remains compressible in the superfluid state and supports longitudinal sound. In fact, experiments with nanomechanical beams in \textsuperscript{4}He have already shown that ballistic phonons and acoustic emission limit the mechanical response at low temperature\cite{Guenault2019He4,Kamppinen2023}. Quartz tuning-fork measurements similarly find that acoustic radiation becomes the dominant damping mechanism at sufficiently high resonator frequency in both \textsuperscript{3}He and \textsuperscript{4}He\cite{Guenault2019He3}. At higher frequencies, nano-optomechanical resonators have been used to demonstrate efficient phonon transfer directly into superfluid \textsuperscript{4}He\cite{Fong2019}. Most of this work has focused on using the mechanics to probe the liquid, or on deliberately hybridizing the solid and fluid acoustic modes. Much less attention has been paid to the inverse question that motivated us here.

To test this idea, we used a cavity magnomechanical resonator, motivated by our attempts to thermalize these devices at millikelvin temperatures.\cite{Huang2025} A yttrium iron garnet (YIG) sphere supports long-lived mechanical modes\cite{LeCraw1961} that couple to magnons through magnetostriction, while the magnons can be strongly coupled to a microwave cavity for readout and control\cite{Zhang2016,Potts2021,Potts2023}. In a previous cryogenic experiment we measured cavity magnomechanics down to \qty{9}{\kelvin}, but further cooling was limited by microwave-power-dependent heating of the YIG sphere\cite{Huang2025}. The basic problem was the same one encountered throughout cryogenic nanomechanics: stronger mechanical support improves thermalization but increases clamping loss, whereas a nearly free sphere has excellent mechanical coherence but a poor thermalization path.

Here, we test superfluid \textsuperscript{4}He as that missing thermal pathway. A \qty{250}{\micro\meter}-diameter YIG sphere is left nearly free inside a partially-sealed glass capillary, and the capillary and microwave cavity are filled with helium. Cavity magnomechanics provides a convenient way to measure both the mechanical linewidth and the temperature-sensitive magnon linewidth of the same sphere. We find that helium immersion strongly damps the mechanical mode. The linewidth of a \qty{12.47}{\mega\hertz} mechanical mode increases by more than three orders of magnitude, and the damping remains large even after the liquid becomes superfluid. Unfortunately, even with this mechanical loss channel, the sphere still shows appreciable microwave heating at the lowest bath temperatures. We show that the mechanical damping is naturally explained by acoustic radiation into the gapless phonon branch of the liquid and that superfluid helium cannot be treated as a lossless exchange medium for compressional mechanical modes.

\begin{figure}
\begin{center}
\includegraphics{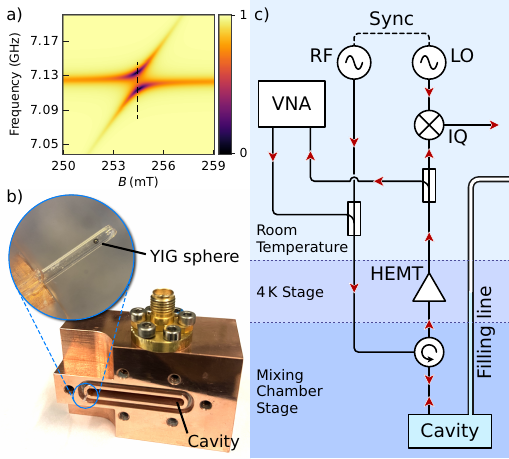}
\end{center}
\caption{\label{fig:setup}(a) Reflection $S_{11}$ versus frequency and magnetic field showing the magnon--photon avoided crossing; the dashed line gives the spectrum in Fig.~\ref{fig:magnonics}. (b) OFHC microwave cavity and, inset, the open glass capsule containing the YIG sphere. (c) Cryogenic microwave measurement and helium fill line.}
\end{figure}


We designed an oxygen-free high-conductivity copper (OFHC) cavity with inner dimensions of \qtyproduct{28 x 28 x 6}{mm}. OFHC has greatly reduced resistivity at the temperatures of this experiment, reducing microwave dissipation from cavity surface currents while maintaining good thermal anchoring to the surroundings. The cavity has a single pin-coupler port on top, similar to the design used in our previous experiments\cite{Potts2020,Huang2025}.
Once the desired pin length is set, this port can be hermetically sealed using low-thermal-contraction epoxy. A second port on the back of the cavity connects to the helium fill line through an indium-sealed bushing, Fig.~1.

Inside the cavity, a \qty{250}{\micro\meter}-diameter YIG sphere is contained in a glass capsule formed from capillary tubing with a \qty{300}{\micro\meter} inner diameter\cite{LeCraw1961,Potts2021}.
Both ends of the capillary are narrowed with a flame, but remain open. The sphere is therefore confined to the capsule while remaining mechanically free over most of its surface, and the open ends allow the capsule to be evacuated and subsequently filled with helium.

\begin{figure}
\begin{center}
\includegraphics{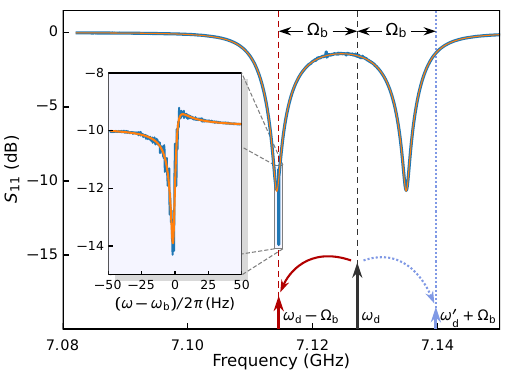}
\end{center}
\caption{\label{fig:magnonics}Hybrid magnon--photon polariton spectrum (blue) and fit (orange). A drive at \(\om d/2\pi=\qty{7.127}{GHz}\) generates sidebands at \(\om d\pm\Om b\), with \(\Om b/2\pi\sim\qty{12.47}{MHz}\). The lower sideband overlaps the lower polariton and is cavity enhanced. The inset shows the measured mechanical feature and Lorentzian fit; the sloped polariton background gives a weak Fano-like asymmetry.}
\end{figure}

The cavity is attached to the mixing-chamber plate of a BlueFors dilution refrigerator with pulse tubes. Microwave signals are input through a cryogenic circulator mounted on the experimental stage before entering the cavity through the pin coupler. The reflected signal is routed by the circulator to the output line of NbTi superconducting coax, and is amplified by a high-electron-mobility-transistor (HEMT) amplifier at the \qty{4}{\kelvin} stage before returning to room temperature. The cavity has a filling port on one side, which connects to a filling capillary through an indium sealed bushing. The filling line has inner diameter of \qty{0.5}{mm}, and is thermally anchored via sintered heat exchangers on each of the \qty{40}K, \qty{4}K, still, cold plate, and mixing chamber plates before reaching the cavity.  When condensing helium into the cavity, helium gas from a canister is first pushed through a nitrogen cold trap, then into the filling line in the refrigerator.

The cavity resonances used for readout are the hybrid polaritons formed by coupling magnons and cavity photons\cite{Flatte2010,Huebl2013,Tabuchi2014,ZhangMagnon2014,Goryachev2014}.
The mechanical motion couples to the magnon mode through magnetostriction\cite{Zhang2016,Potts2021,Potts2023}. When measured in reflection, the polariton spectrum is
\begin{equation}\label{eq:spectrum}
    S_{11}(\omega) = 1 - \frac{\mathrm{e}^{i\eta_\textrm{mis}}\kappa_\textrm{ext}}{i(\omega - \omega_a) + \frac{\kappa_\textrm{int} + \kappa_\textrm{ext}}{2} + \frac{g_\textrm{am}^2}{i(\omega - \omega_m) + \gamma_\textrm{m}/2} },
\end{equation}
where \(\kappa_\textrm{int}\) and \(\kappa_\textrm{ext}\) are the internal and external cavity loss rates, \(\gamma_\textrm{m}\) is the magnon linewidth, and \(\eta_\textrm{mis}\) accounts for impedance mismatch. 
A VNA sweep of \(S_{11}\) therefore allows us to extract the cavity, magnon, and coupling parameters by fitting Eq.~\ref{eq:spectrum}.

A second microwave tone at frequency \(\om d\) drives the magnomechanical interaction and produces sidebands at \(\om d\pm\Om b\). When one sideband overlaps a polariton resonance, the mechanical scattering is cavity enhanced and interferes with the weak VNA probe. This produces magnomechanically induced transparency or absorption (MMIT/MMIA)\cite{Zhang2016,Potts2021,Potts2023}, similar to optomechanically induced transparency (OMIT)\cite{Weis2010} or electromagnetically induced transparency (EIT)\cite{Fleischhauer2005}.
The orange trace in Fig.~\ref{fig:magnonics} is the cavity enhanced \(\om d-\Om b\) scattering sideband. Fitting this narrow mechanical feature gives the mechanical resonance frequency and linewidth.

\begin{figure}
\begin{center}
\includegraphics{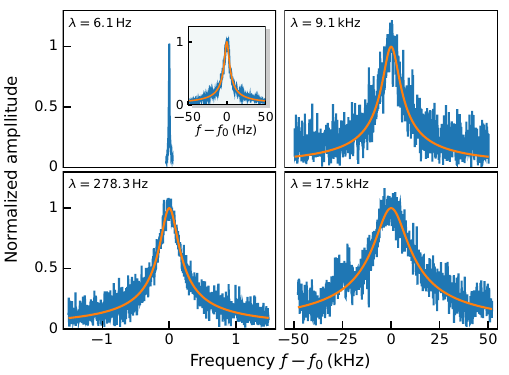}
\end{center}
\caption{\label{fig:mech_lw}Mechanical spectra (blue) and Lorentzian fits (orange), normalized and centered for comparison. Conditions are evacuated at \(T\sim\qty{4.2}{K}\) (top left), partial fill at \qty{4.2}{K} (bottom left), full fill at \qty{4.2}{K} (top right), and full fill at \qty{84}{mK} (bottom right). The fitted linewidths are \qty{6.1}{Hz}, \qty{278}{Hz}, \qty{9.1}{kHz}, and \qty{17.5}{kHz}, respectively.}
\end{figure}

We measured polaritons and mechanics at various temperatures. At each temperature, we stabilized the mixing chamber temperature using a PID loop controlling a heater on the stage. At temperatures closer to the base temperature ($\sim \qty{15}{mK}$), the reduced cooling power of the dilution unit causes the temperature to become less stable at higher microwave powers, therefore the results presented here were from slightly higher temperatures (see Fig.~\ref{fig:mech_lw} for example), nonetheless deep into the superfluid state.


We first cooled the cryostat to approximately \qty{4.2}{\kelvin} with the cavity evacuated and measured the polariton and MMIT spectra. Helium was then condensed slowly into the cavity while the mechanical response was monitored. The progressive broadening of the mechanical resonance provided a direct indication that liquid helium was filling the capsule. After the cavity had been filled overnight, we ran the dilution refrigerator and cooled the filled-cavity to the mK regime.

Representative mechanical spectra are shown in Fig.~\ref{fig:mech_lw}. Before condensation, the \(\sim\qty{12.47}{\mega\hertz}\) mechanical mode has a linewidth of only \qty{5.8}{\hertz}, corresponding to \(Q\approx2.1\times10^6\). Part way through the fill the linewidth has already increased to \qty{316}{\hertz}. Once the cavity is completely filled with normal liquid helium at \qty{4.2}{\kelvin}, the linewidth is \qty{8.0}{\kilo\hertz}, corresponding to \(Q\approx1.6\times10^3\). Cooling through the superfluid transition does not recover the narrow mechanical resonance. At a mixing-chamber temperature of \qty{84}{\milli\kelvin}, the linewidth is instead \qty{18.5}{\kilo\hertz}, or \(Q\approx6.7\times10^2\). Thus immersion in helium reduces the mechanical quality factor by more than three orders of magnitude. 

\begin{figure}
\begin{center}
\includegraphics{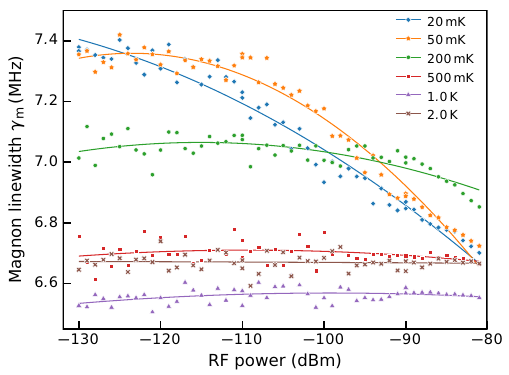}
\end{center}
\caption{\label{fig:gamma_m_vs_P}YIG magnon linewidth \(\gamma_\textrm{m}\) versus calibrated microwave power delivered to the cavity. The VNA probe is kept much weaker than the RF drive. Strong power dependence at the lowest temperatures indicates microwave heating even in superfluid helium; lines are guides to the eye.}
\end{figure}

The large mechanical damping answers one part of the question, but not whether the YIG sphere itself is well thermalized. For this we use the magnon linewidth as an \textit{in situ} thermometer. In our previous cryogenic measurements, the YIG magnon linewidth showed a reproducible temperature dependence and provided a sensitive probe of microwave heating of the sphere\cite{Huang2025}. Figure~\ref{fig:gamma_m_vs_P} shows the magnon linewidth measured here as a function of microwave drive power at several mixing-chamber temperatures.

At the higher bath temperatures the magnon linewidth changes only weakly with applied microwave power. At the lowest temperatures, however, the linewidth changes strongly as the drive is increased. 
When starting at the lowest temperatures, but increasing the microwave drive power, the magnon linewidth converges towards the values at higher temperatures, suggesting heating to hundreds of millikelvin.  
We therefore do not find evidence that immersion in superfluid helium effectively thermalizes the YIG sphere to the mixing chamber temperature under localized heat load.  This also reiterates the value of having an \textit{in situ} temperature sensor, such as the magnon linewidth.  


\begin{figure}
\begin{center}
\includegraphics{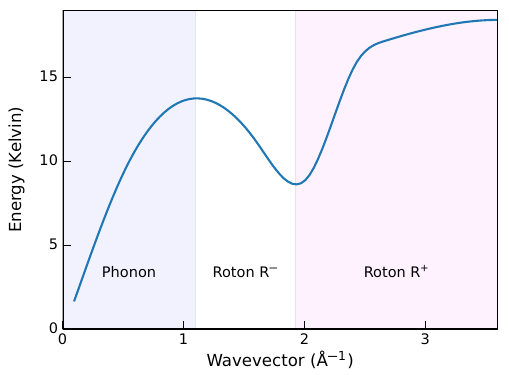}
\end{center}
\caption{\label{fig:helium_dispersion}Quasiparticle dispersion of liquid \textsuperscript{4}He. The gapless phonon branch provides the longitudinal acoustic channel relevant to the YIG mechanical mode.}
\end{figure}

In hindsight, the persistence of large damping in the superfluid should not be surprising. Superfluidity removes viscous drag associated with the superfluid component, but it does not make the liquid incompressible or change its acoustic modes. The YIG mechanical mode used here has predominantly normal displacement at the sphere surface\cite{Potts2021}. That motion is therefore well matched to longitudinal pressure waves in the surrounding liquid. At low temperature the relevant excitations of superfluid \textsuperscript{4}He lie on the gapless phonon branch of the dispersion relation\cite{Landau1941,Donnelly1998}, shown schematically in Fig.~\ref{fig:helium_dispersion}. Mechanical energy can therefore be radiated continuously from the YIG into helium as first sound even when viscous damping has become negligible. This is the same basic acoustic-emission mechanism identified in high-frequency nanobeams and tuning forks immersed in helium\cite{Guenault2019He4,Guenault2019He3}, and is consistent with the efficient solid-to-superfluid phonon coupling observed in nano-optomechanical devices\cite{Fong2019}.

The scale of the observed linewidth can be understood with a simple estimate. At \(f_m\approx\qty{12.47}{\mega\hertz}\), the low-temperature sound speed of liquid \textsuperscript{4}He, \(c\approx\qty{238}{\meter\per\second}\), gives an acoustic wavelength of \(\lambda\approx\qty{19}{\micro\meter}\)\cite{Donnelly1998}. For a sphere of radius \(a=\qty{125}{\micro\meter}\), \(ka=2\pi a/\lambda\approx41\). The surface is therefore large compared with an acoustic wavelength, and a normally moving surface element sees a radiation resistance approaching the plane-wave acoustic impedance, \(\rho_\mathrm{He}c\). The time-averaged radiated power is then approximately
\begin{equation}
P_\mathrm{ac}\sim \frac{1}{2}\rho_\mathrm{He}c\int |v_n|^2\,dA,
\label{eq:pac}
\end{equation}
where \(v_n\) is the normal surface velocity. Using the entire sphere surface and the full sphere mass gives
\begin{equation}
\Delta f_\mathrm{ac}\sim
\frac{\rho_\mathrm{He}c\,4\pi a^2}{2\pi m_\mathrm{sph}}
\approx \qty{25}{\kilo\hertz},
\label{eq:acousticlinewidth}
\end{equation}
with \(\rho_\mathrm{He}\approx\qty{145}{\kilogram\per\meter\cubed}\) and \(m_\mathrm{sph}\approx\qty{42}{\micro\gram}\). While this is only an order-of-magnitude estimate, comparison with the measured \qty{17.5}{\kilo\hertz} linewidth, the acoustic power lost from the surface of the sphere dictates the mechanical dissipation. At this frequency, superfluid helium is a lossy acoustic medium.

This is not specific to cavity magnomechanics. If a mechanical mode moves a surface normal to a surrounding liquid, the same coupling that makes the liquid an attractive thermal reservoir also makes it an acoustic reservoir.  For MHz mechanical resonators with dimensions larger than the sound wavelength, the acoustic load can be large. 

There is, however, a possible way around this. A mechanical mode dominated by transverse shear motion should couple much more weakly to bulk compression waves in an ordinary liquid, because a liquid does not support static shear stress or propagating transverse sound. Shear-polarized resonators are therefore more promising candidates with the goal to use superfluid \textsuperscript{4}He as a thermal exchange medium, while preserving a high mechanical quality factor.  This is not true in \textsuperscript{3}He though, where transverse acoustic modes in superfluid \textsuperscript{3}He-B propagate via coupling to order-parameter collective modes\cite{Moores1993,Lee1999,Zheng2016,Zheng2017,Jiang2023, Davis2008a, Davis2008b}.


We immersed a nearly free YIG mechanical resonator in liquid \textsuperscript{4}He to test a simple idea that has circulated in the cryogenic mechanics community: use superfluid helium as an exchange medium to improve thermalization without paying the usual viscous damping penalty. For the predominantly normal-displacement mode studied here, this does not work. The mechanical linewidth increases from \qty{5.8}{\hertz} in vacuum to \qty{18.5}{\kilo\hertz} in the low-temperature superfluid, reducing the quality factor by more than three orders of magnitude. Meanwhile, the YIG magnon linewidth remains power dependent at the lowest temperatures, showing that significant microwave heating of the sphere can still occur.

The origin of the damping is straightforward. The superfluid component is inviscid, but superfluid \textsuperscript{4}He still supports gapless longitudinal sound. A \qty{12.47}{\mega\hertz} mode of a \qty{250}{\micro\meter} sphere is therefore an efficient acoustic radiator, and a simple radiation-impedance estimate produces a linewidth on the same scale as the experiment. This conclusion is not specific to YIG or to cavity magnomechanics: superfluid helium should not be assumed to be a low-loss thermal bath for a mechanical mode simply because it is a superfluid.

A possible way forward is to separate the thermal and mechanical coupling by geometry. A shear mode with little normal surface displacement should radiate far less first sound into \textsuperscript{4}He while retaining direct contact with the liquid for heat transport. Testing such modes would determine whether superfluid helium can still serve as a useful cryogenic exchange medium when the dominant longitudinal acoustic channel is engineered away.

\bigskip

The authors acknowledge that the land on which this work was performed is on Treaty Six Territory, the traditional lands of many First Nations, Métis, and Inuit in Alberta. They acknowledge the support from the Natural Sciences and Engineering Research Council, Canada (Grant Nos.~RGPIN-2022-03078,  ALLRP 592535-23, ALLRP 568609-21); Alberta Innovates (Grant No.~212200780); the National Research Council (Grant QSP-017-1); and the University of Alberta Department of Physics through the Advah Bhatia Fellowship.


\begin{thebibliography}{99}

\bibitem{Aspelmeyer2014}
M. Aspelmeyer, T. J. Kippenberg, and F. Marquardt, ``Cavity optomechanics,'' Rev. Mod. Phys. \textbf{86}, 1391--1452 (2014).

\bibitem{OConnell2010}
A. D. O'Connell \textit{et al.}, ``Quantum ground state and single-phonon control of a mechanical resonator,'' Nature \textbf{464}, 697--703 (2010).

\bibitem{Teufel2011}
J. D. Teufel \textit{et al.}, ``Sideband cooling of micromechanical motion to the quantum ground state,'' Nature \textbf{475}, 359--363 (2011).

\bibitem{Reinhardt2016}
C. Reinhardt, T. M\"uller, A. Bourassa, and J. C. Sankey, ``Ultralow-noise SiN trampoline resonators for sensing and optomechanics,'' Phys. Rev. X \textbf{6}, 021001 (2016).

\bibitem{Norte2016}
R. A. Norte, J. P. Moura, and S. Gr\"oblacher, ``Mechanical resonators for quantum optomechanics experiments at room temperature,'' Phys. Rev. Lett. \textbf{116}, 147202 (2016).

\bibitem{Tsaturyan2017}
Y. Tsaturyan, A. Barg, E. S. Polzik, and A. Schliesser, ``Ultracoherent nanomechanical resonators via soft clamping and dissipation dilution,'' Nat. Nanotechnol. \textbf{12}, 776--783 (2017).

\bibitem{Fink2016}
J. M. Fink, M. Kalaee, A. Pitanti, R. Norte, L. Heinzle, M. Davan\c{c}o, K. Srinivasan, and O. Painter, ``Quantum electromechanics on silicon nitride nanomembranes,'' Nat. Commun. \textbf{7}, 12396 (2016).

\bibitem{Ramp2019}
H. Ramp, B. D. Hauer, K. C. Balram, T. J. Clark, K. Srinivasan, and J. P. Davis, ``Elimination of thermomechanical noise in piezoelectric optomechanical crystals,'' Phys. Rev. Lett. \textbf{123}, 093603 (2019).

\bibitem{MacCabe2020}
G. S. MacCabe, H. Ren, J. Luo, J. D. Cohen, H. Zhou, A. Sipahigil, M. Mirhosseini, and O. Painter, ``Nano-acoustic resonator with ultralong phonon lifetime,'' Science \textbf{370}, 840--843 (2020).

\bibitem{Meenehan2014}
S. M. Meenehan, J. D. Cohen, S. Gr\"oblacher, J. T. Hill, A. H. Safavi-Naeini, M. Aspelmeyer, and O. Painter, ``Silicon optomechanical crystal resonator at millikelvin temperatures,'' Phys. Rev. A \textbf{90}, 011803(R) (2014).

\bibitem{Meenehan2015}
S. M. Meenehan, J. D. Cohen, G. S. MacCabe, F. Marsili, M. D. Shaw, and O. Painter, ``Pulsed excitation dynamics of an optomechanical crystal resonator near its quantum ground state of motion,'' Phys. Rev. X \textbf{5}, 041002 (2015).

\bibitem{MacDonald2016}
A. J. R. MacDonald, B. D. Hauer, X. Rojas, P. H. Kim, G. G. Popowich, and J. P. Davis, ``Optomechanics and thermometry of cryogenic silica microresonators,'' Phys. Rev. A \textbf{93}, 013836 (2016).

\bibitem{Kim2016}
P. H. Kim, B. D. Hauer, C. Doolin, F. Souris, and J. P. Davis, ``Approaching the standard quantum limit of mechanical torque sensing,'' Nat. Commun. \textbf{7}, 13165 (2016).

\bibitem{Hauer2018}
B. D. Hauer, P. H. Kim, C. Doolin, F. Souris, and J. P. Davis, ``Two-level system damping in a quasi-one-dimensional optomechanical resonator,'' Phys. Rev. B \textbf{98}, 214303 (2018).

\bibitem{Bradley2017}
D. I. Bradley \textit{et al.}, ``Operating nanobeams in a quantum fluid,'' Sci. Rep. \textbf{7}, 4876 (2017).

\bibitem{Lane2020}
J. R. Lane, D. Tan, N. R. Beysengulov, K. Nasyedkin, E. Brook, L. Zhang, T. Stefanski, H. Byeon, K. W. Murch, and J. Pollanen, ``Integrating superfluids with superconducting qubit systems,'' Phys. Rev. A \textbf{101}, 012336 (2020).

\bibitem{Li2025}
C. Li, Y. Wu, M. C. C. Pace, M. D. LaHaye, M. Senatore, and H. X. Tang, ``Superfluid-cooled transmon qubits under optical excitation,'' PRX Quantum \textbf{6}, 030303 (2025).

\bibitem{Guenault2019He4}
A. M. Gu\'enault, A. Guthrie, R. P. Haley, S. Kafanov, Yu. A. Pashkin, G. R. Pickett, M. Poole, R. Schanen, V. Tsepelin, D. E. Zmeev, E. Collin, O. Maillet, and R. Gazizulin, ``Probing superfluid \textsuperscript{4}He with high-frequency nanomechanical resonators down to millikelvin temperatures,'' Phys. Rev. B \textbf{100}, 020506(R) (2019).

\bibitem{Barquist2020}
C. S. Barquist, W. G. Jiang, K. Gunther, N. Eng, Y. Lee, and H. B. Chan, ``Damping of a microelectromechanical oscillator in turbulent superfluid \textsuperscript{4}He: A probe of quantized vorticity in the ultralow temperature regime,'' Phys. Rev. B \textbf{101}, 174513 (2020).

\bibitem{Guenault2019He3}
A. M. Gu\'enault, R. P. Haley, S. Kafanov, M. T. Noble, G. R. Pickett, M. Poole, R. Schanen, V. Tsepelin, J. Vonka, T. Wilcox, and D. E. Zmeev, ``Acoustic damping of quartz tuning forks in normal and superfluid \textsuperscript{3}He,'' Phys. Rev. B \textbf{100}, 104526 (2019).

\bibitem{Zheng2016}
P. Zheng, W. G. Jiang, C. S. Barquist, Y. Lee, and H. B. Chan, ``Anomalous damping of a microelectromechanical oscillator in superfluid \textsuperscript{3}He-B,'' Phys. Rev. Lett. \textbf{117}, 195301 (2016).

\bibitem{Zheng2017}
P. Zheng, W. G. Jiang, C. S. Barquist, Y. Lee, and H. B. Chan, ``Critical velocity in the presence of surface bound states in superfluid \textsuperscript{3}He-B,'' Phys. Rev. Lett. \textbf{118}, 065301 (2017).

\bibitem{Jiang2023}
W. G. Jiang, C. S. Barquist, K. Gunther, Y. Lee, and H. B. Chan, ``Superfluid \textsuperscript{3}He-B surface states in a confined geometry probed by a microelectromechanical oscillator,'' Phys. Rev. B \textbf{108}, 024508 (2023).

\bibitem{Rojas2015}
X. Rojas and J. P. Davis, ``Superfluid nanomechanical resonator for quantum nanofluidics,'' Phys. Rev. B \textbf{91}, 024503 (2015).

\bibitem{Souris2017}
F. Souris, X. Rojas, P. H. Kim, and J. P. Davis, ``Ultralow-dissipation superfluid micromechanical resonator,'' Phys. Rev. Applied \textbf{7}, 044008 (2017).

\bibitem{Baker2016}
C. G. Baker, G. I. Harris, D. L. McAuslan, Y. Sachkou, X. He, and W. P. Bowen, ``Theoretical framework for thin film superfluid optomechanics: Towards the quantum regime,'' New J. Phys. \textbf{18}, 123025 (2016).

\bibitem{Kashkanova2017}
A. D. Kashkanova, A. B. Shkarin, C. D. Brown, N. E. Flowers-Jacobs, L. Childress, S. W. Hoch, L. Hohmann, K. Ott, J. Reichel, and J. G. E. Harris, ``Superfluid Brillouin optomechanics,'' Nat. Phys. \textbf{13}, 74--79 (2017).

\bibitem{Shkarin2019}
A. B. Shkarin, A. D. Kashkanova, C. D. Brown, S. Garcia, K. Ott, J. Reichel, and J. G. E. Harris, ``Quantum optomechanics in a liquid,'' Phys. Rev. Lett. \textbf{122}, 153601 (2019).

\bibitem{Kamppinen2023}
T. Kamppinen, J. T. M\"akinen, and V. B. Eltsov, ``Superfluid \textsuperscript{4}He as a rigorous test bench for different damping models in nanoelectromechanical resonators,'' Phys. Rev. B \textbf{107}, 014502 (2023).

\bibitem{Fong2019}
K. Y. Fong, D. Jin, M. Poot, A. Bruch, and H. X. Tang, ``Phonon coupling between a nanomechanical resonator and a quantum fluid,'' Nano Lett. \textbf{19}, 3716--3722 (2019).

\bibitem{Huang2025}
Y. Huang, P. M. C. Rourke, A. Peruzzi, J. Jin, M. Ebrahimi, A. Rashedi, and J. P. Davis, ``Cryogenic magnomechanics for thermometry applications,'' Appl. Phys. Lett. \textbf{127}, 082405 (2025).

\bibitem{LeCraw1961}
R. C. LeCraw, E. G. Spencer, and E. I. Gordon, ``Extremely
Low Loss Acoustic Resonance in Single-Crystal Garnet
Spheres'', Phys. Rev. Lett. \textbf{6}, 620 (1961).

\bibitem{Zhang2016}
X. Zhang, C.-L. Zou, L. Jiang, and H. X. Tang, ``Cavity magnomechanics,'' Sci. Adv. \textbf{2}, e1501286 (2016).

\bibitem{Potts2021}
C. A. Potts, E. Varga, V. A. S. V. Bittencourt, S. Viola Kusminskiy, and J. P. Davis, ``Dynamical backaction magnomechanics,'' Phys. Rev. X \textbf{11}, 031053 (2021).

\bibitem{Potts2023}
C. A. Potts, Y. Huang, V. A. S. V. Bittencourt, S. Viola Kusminskiy, and J. P. Davis, ``Dynamical backaction evading magnomechanics,'' Phys. Rev. B \textbf{107}, L140405 (2023).

\bibitem{Potts2020}
C. A. Potts and J. P. Davis, ``Strong magnon--photon coupling within a tunable cryogenic microwave cavity,'' Appl. Phys. Lett. \textbf{116}, 263503 (2020).

\bibitem{Flatte2010}
O. O. Soykal and M. E. Flatt\'e, ``Strong Field Interactions between a Nanomagnet and a Photonic Cavity'', Phys. Rev. Lett. \textbf{104}, 077202 (2010).

\bibitem{Huebl2013}
H. Huebl, C. W. Zollitsch, J. Lotze, F. Hocke, M. Greifenstein, A. Marx, R. Gross, S. T. B Goennenwein, ``High Cooperativity in Coupled Microwave Resonator Ferrimagnetic Insulator Hybrids'', Phys. Rev. Lett. \textbf{111}, 127003 (2013).

\bibitem{Tabuchi2014}
Y. Tabuchi, S. Ishino, T. Ishikawa, R. Yamazaki, K. Usami, and Y. Nakamura, ``Hybridizing ferromagnetic magnons and microwave photons in the quantum limit,'' Phys. Rev. Lett. \textbf{113}, 083603 (2014).

\bibitem{ZhangMagnon2014}
X. Zhang, C.-L. Zou, L. Jiang, and H. X. Tang, ``Strongly coupled magnons and cavity microwave photons,'' Phys. Rev. Lett. \textbf{113}, 156401 (2014).

\bibitem{Goryachev2014}
M. Goryachev, W. G. Farr, D. L. Creedon, Y. Fan, M. Kostylev, and M. E. Tobar, ``High-cooperativity cavity QED with magnons at microwave frequencies,'' Phys. Rev. Applied \textbf{2}, 054002 (2014).

\bibitem{Weis2010}
S. Weis, R. Rivi\`ere, S. Del\'eglise, E. Gavartin, O. Arcizet, A. Schliesser, and T. J. Kippenberg, ``Optomechanically induced transparency,'' Science \textbf{330}, 1520--1523 (2010).

\bibitem{Fleischhauer2005}
M. Fleischhauer, A. Imamoglu, J. P. Marangos, ``Electromagnetically induced transparency: Optics in coherent media'',  Rev. Mod. Phys. \textbf{77}, 633--673 (2005).

\bibitem{Swartz1989}
E. T. Swartz and R. O. Pohl, ``Thermal boundary resistance,'' Rev. Mod. Phys. \textbf{61}, 605--668 (1989).

\bibitem{Landau1941}
L. Landau, ``Theory of the superfluidity of helium II,'' Phys. Rev. \textbf{60}, 356--358 (1941).

\bibitem{Donnelly1998}
R. J. Donnelly and C. F. Barenghi, ``The observed properties of liquid helium at the saturated vapor pressure,'' J. Phys. Chem. Ref. Data \textbf{27}, 1217--1274 (1998).

\bibitem{Moores1993}
G. F. Moores and J. A. Sauls, ``Transverse waves in superfluid \textsuperscript{3}He-B,'' J. Low Temp. Phys. \textbf{91}, 13--37 (1993).

\bibitem{Lee1999}
Y. Lee, T. M. Haard, W. P. Halperin, and J. A. Sauls, ``Discovery of the acoustic Faraday effect in superfluid \textsuperscript{3}He-B,'' Nature \textbf{400}, 431--433 (1999).

\bibitem{Davis2008a}
J. P. Davis, J. Pollanen, H. Choi, J. A. Sauls, and W. P. Halperin, ``Discovery of an excited pair state in superfluid \textsuperscript{3}He,''
Nature Physics \textbf{4}, 571-–575 (2008)

\bibitem{Davis2008b}
J. P. Davis, J. Pollanen, H. Choi, J. A. Sauls,  W. P. Halperin, A. B. Vorontsov, ``Anomalous Attenuation of Transverse Sound in \textsuperscript{3}He,'' Phys. Rev. Lett. 101, 085301 (2008).




\end{thebibliography}
\end{document}